# Recent Advances in High-Temperature Superconductivity

Nai-Chang Yeh

*After more than 15 years of intense research since the discovery of high-temperature superconductivity [1], many interesting physical phenomena unique to the cuprate superconductors are better understood, and various applications have been realized. However, the underlying mechanism for high-temperature superconductivity remains elusive, largely due to the complication of numerous competing orders in the ground state of the cuprates. We review some of the most important physics issues and recent experimental developments associated with these strongly correlated electronic systems, and discuss current understanding and possible future research direction.*

## 1. INTRODUCTION

High-temperature superconducting cuprates are doped Mott insulators with numerous competing orders in the ground state [2-5]. Mott insulators differ from conventional band insulators in that the latter are dictated by the Pauli exclusion principle when the highest occupied band contains two electrons per unit cell, whereas the former are associated with the existence of strong on-site Coulomb repulsion such that double occupancy of electrons per unit cell is energetically unfavorable and the electronic system behaves like an insulator rather than a good conductor at half filling. An important signature of doped Mott insulators is the strong electronic correlation among the carriers and the sensitivity of their ground state to the doping level. In cuprates, the ground state of the undoped perovskite oxide is an antiferromagnetic Mott insulator, with nearest-neighbor $Cu^{2+}$-$Cu^{2+}$ antiferromagnetic exchange interaction in the $CuO_2$ planes [6]. Depending on doping with either electrons or holes into the $CuO_2$ planes [6,7], the Néel temperature ($T_N$) for the antiferromagnetic-to-paramagnetic transition decreases with increasing doping level. Upon further doping of carriers, long-range antiferromagnetism vanishes and is replaced by superconductivity. As shown in the phase diagrams for the hole-doped (p-type) and electron-doped (n-type) cuprates in Fig. 1, the superconducting transition temperature ($T_c$) first increases with increasing doping level ($\delta$), reaching a maximum $T_c$ at an optimal doping level, then decreases and finally vanishes with further increase of doping. Although the phase diagrams appear similar for both p-type and n-type cuprates, they are in fact not truly symmetric. For *p*-type cuprates in the under- and optimally doped regime, the normal state properties below a crossover temperature $T^*$ are significantly different from those of Fermi liquid, and the electronic density of states (DOS) appear to be slightly suppressed [8]. These unconventional normal state properties are referred to as the pseudogap phenomenon [8]. Moreover, holes enter

Nai-Chang Yeh
Professor of Physics
California Institute of Technology
Pasadena, CA 91125, USA
E-mail: ncyeh@caltech.edu

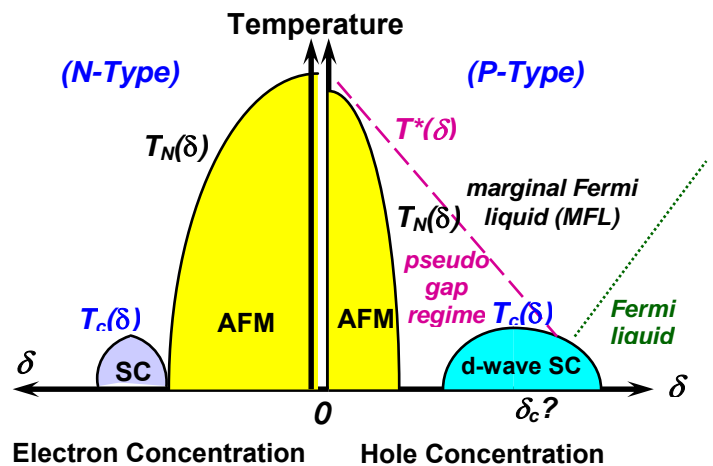

Fig. 1: *Generic temperature (T) vs. doping level (δ) phase diagrams of p-type and n-type cuprates in zero magnetic field. (AFM: antiferromagnetic phase; SC: superconducting phase; $T_N$, $T_c$ and $T^*$ are the Néel, superconducting and pseudogap transition temperatures, respectively).*

into the oxygen *p*-orbital in the CuO$_2$ planes, which induce ferromagnetic coupling for the Cu$^{2+}$ ions adjacent to the partially empty oxygen orbital, thus resulting in significant spin frustrations in the CuO$_2$ planes, as schematically illustrated in Fig. 2(a) for a specific 1/8-doping level. The resulting strong spin fluctuations are the primary cause for the rapid decline of the Néel state with increasing hole doping. On the other hand, electron doping in n-type cuprates takes place in the *d*-orbital of Cu, giving rise to spinless Cu$^+$-ions that dilute the background antiferromagnetic Cu$^{2+}$-Cu$^{2+}$ coupling without inducing as strong spin frustrations as those in the p-type cuprates, as shown in Figure 2(b). Hence, the Néel state survives over a larger range of electron doping, in contrast to the p-type cuprates, whereas the superconducting phase in the n-type cuprates exists over a much narrower doping range relative to the p-type cuprates. Other important contrasts between the n-type and p-type cuprates include the absence of pseudogap phenomena in the former [9-12], non-universal pairing symmetries [9-22], and different doping-dependent Fermi surface evolution according to the angular-resolved photoemission spectroscopy (ARPES) [23]. The lack of electron-hole symmetry suggests that the cuprates cannot be fully described by a one-band Hubbard or *t-J* model.

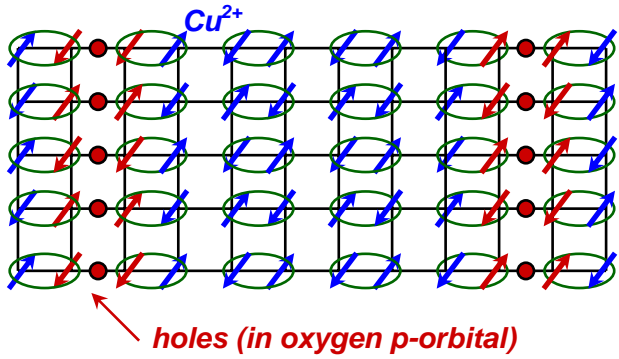

**(a) CuO$_2$ plane (for δ ~ 1/8 hole doping)**

holes (in oxygen p-orbital)

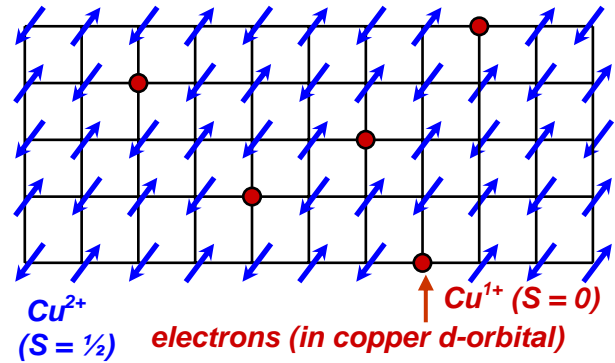

**(b) CuO$_2$ plane (for δ ~ 0.1 electron doping)**

Cu$^{2+}$ (S = ½)   electrons (in copper d-orbital)   Cu$^{1+}$ (S = 0)

*Fig. 2: Effects of hole and electron doping on the spin configurations in the CuO$_2$ plane. (a) Doped holes are associated with the oxygen p-orbital, which result in ferromagnetic coupling between the neighboring Cu$^{2+}$ ions, yielding strong spin frustrations in the CuO$_2$ plane. For a special 1/8 doping level, charge stripes can be formed as illustrated. (b) Doped electrons are associated with the d-orbital of Cu, yielding Cu$^+$ that dilutes the antiferromagnetism of the undoped sample without causing significant spin frustrations.*

Concerning the competing orders in the ground state of the cuprates, besides the obvious SU(2) and U(1) broken symmetries associated with the occurrence of antiferromagnetism and superconductivity, respectively, other competing orders include the crystalline symmetry (C) and the time-reversal (T) symmetry [3,4]. These competing orders in the two-dimensional one-band square-lattice approximation can give rise to a large variety of doping-dependent ground states [3,4]. For instance, charge stripes can exist under specific doping levels (e.g. 1/8), as exemplified in Fig. 2(a), which have observed in some underdoped cuprates [24-28]. Another possible ground state is the *d*-density wave (DDW) state also known as orbital antiferromagnetism [29], which involves alternating orbital currents from one plaque to the adjacent plaque [29,30]. The DDW state is a broken T-symmetry state, which in principle can be verified experimentally [29], although to date no conclusive empirical evidence has been found. Other possible ground states based on the simplified mean-field and two-dimensional square-lattice approximations include the spin-Pierrl state, Wigner crystal state, spin density waves (SDW), charge density waves (CDW), and complex pairing symmetry of ($d_{x^2-y^2}+id_{xy}$) or ($d_{x^2-y^2}+is$), depending on the doping level and the Coulomb and exchange interaction strengths [3,4]. The large varieties of ground states are indicative of the complex nature of competing orders in the cuprates. It is therefore imperative to identify universal characteristics among all cuprates and to develop understanding for the differences in order to unravel the underlying pairing mechanism for cuprate superconductivity.

## 2. THEORETICAL CONJECTURES AND COMPARISON WITH EXPERIMENTS

There has been no consensus to date for the mechanism of cuprate superconductivity, and the theoretical status of the field has been largely phenomenological and controversial. In general, the ground state of cuprates depends sensitively on the doping level, the type of carriers, the electronic coupling between adjacent $CuO_2$ layers, and the degree of disorder. The complication of competing orders [2-5] and the resulting rich experimental phenomena are the primary contributors to the lack of theoretical consensus. In this section, we review representative theoretical scenarios and compare them with available experimental results. Some of the important recent experimental developments will be discussed in more details in the next section.

### 2.1. Conjectures for the Pairing State

One of the earliest theoretical conjectures for cuprate superconductivity is the resonating valence bond (RVB) theory for p-type cuprates [31-33]. An important consequence of the RVB theory is spin-charge separation, so that the low-energy excitations consist of spinons and holons rather than quasiparticles, and the normal state properties differ from the Fermi liquid behavior for typical metals. The existence of an RVB phase would require strong quantum fluctuations so that the ground state of the undoped sample is a quantum liquid rather than a long-range ordered Néel state. However, it has been unambiguously verified that the ground state of the strongly underdoped cuprates is a well-defined Néel state [6], and no direct evidence for spin-charge separation has been obtained [34,35]. A related model inspired by the RVB theory is the interlayer pair tunneling (ILPT) scenario [36], which asserts that the unusual normal state of the cuprates prohibits coherent single particle transport between the $CuO_2$ planes, whereas pair tunneling becomes possible in the superconducting state. Thus, cuprate superconductivity could arise from interlayer pairing tunneling and the condensation energy would be directly proportional to the kinetic energy of pair tunneling [36,37]. However, irreconcilable discrepancies have been found between explicit predictions of the ILPT scenario [36,37] and experimental results [38]. Hence, both the RVB and ILPT conjectures appear to be inconsistent with empirical facts.

Another important theoretical model is the spin fluctuation scenario [39,40], which suggests that superconductivity arises in the $CuO_2$ planes and is strongly related to the in-plane AFM correlation. An explicit prediction of the spin fluctuation scenario is the $d_{x^2-y^2}$-wave pairing symmetry. While compatible with most experimental phenomena associated with p-type cuprates, recent findings of s-wave pairing and the absence of gapped spin fluctuations in n-type cuprate superconductors [9,10,17,20] have impose difficulties on this model.

In general, most theories for cuprate superconductivity have based on the assumption that the pseudogap phenomenon is a precursor for cuprate superconductivity [3-5]. These theories may be tentatively divided into two categories. One category associates the onset of Cooper pairing with the establishment of AFM coupling of nearest-neighbor $Cu^{2+}$ ions. Thus, the effective mean-field transition temperature would be of the order of the magnetic coupling energy $J$, which is between 250 K and 400 K, much larger than the superconducting condensation energy [41]. The other category centers on strong fluctuation effects of the superconducting order parameter due to the small phase stiffness of the cuprates [42,43]. Representative models associated with the latter concept include the conjectures of Josephson coupling of charged stripes at $T_c$ [44,45], Bose-Einstein condensation (BEC) of preformed Cooper pairs [46,47] at $T_c$, and the vison hypothesis [48]. Other models assuming strong quantum fluctuations include the RVB theory described earlier, the conjecture of a quantum critical point (QCP) near the optimal doping level [49] and the accompanying circulating current phase [49,50]. On the other hand, models associated with the conjecture of magnetic pairing include the SU(2) slave-boson scenario [51,52], the SO(5) quantum non-linear σ-model [5,53], and mean-field consideration of competing orders in the ground state as a function of the doping level and varying strengths of the Coulomb interaction [4].

The stripe scenario asserts that charged stripes are generic in all cuprates and that preformed Cooper pairs exist in the pseudogap regime (i.e., at $T_c < T < T^*$) because of strong phase fluctuations [42]. Global superconductivity becomes established when the carrier concentration is above the percolation threshold so that Josephson coupling among stripes can be established to achieve global phase coherence at $T < T_c$ [44,45]. In addition, spin-charge separation can be expected in the quasi-one dimensional charged stripes in the pseudogap regime. However, among all families of cuprates, the majority exhibits incompatibility of superconductivity with static stripes. The conjecture that dynamic stripes may exist and oscillate at frequencies higher than most experimental probes also seems unphysical because rapidly fluctuating charged stripes would have resulted in significant radiation, which is obviously inconsistent with experimental observation. Recent neutron scattering experiments on strongly underdoped p-type cuprate superconductors (such as $La_{1.6-x}Nd_{0.4}Sr_xCuO_4$ [25], $La_{1.875}Ba_{0.125-x}Sr_xCuO_4$ [26] and $YBa_2Cu_3O_{6.35}$ [27]) have found that static charged stripes can coexist with superconductivity if the spin order remains dynamic. These experimental findings are consistent with charged stripes being a consequence of competing orders rather than a sufficient and ubiquitous condition for cuprate superconductivity.

Despite difficulties associated with the stripe scenario, the tendency of cuprates to forming short-range charge stripes can actually account for the gapped incommensurate spin fluctuations associated with p-type cuprates, as observed in neutron scattering experiments [6,28]. That is, incommensurate spin fluctuations in p-type cuprates are found to correlate with the charge doping level $\delta$, so that in the underdoped regime, spin excitations occur at $Q_\delta = [½, ½(1 \pm \delta)]$ and $[½(1 \pm \delta), ½]$. These spin excitations may be understood in terms of the stripe scenario illustrated in Fig. 3. That is, the spin configurations can be locally commensurate and are adjusted to the doping level by abrupt jump of a phase $\pi$ at periodic charge stripes that serve as antiphase boundaries for the spins [2,44], as illustrated in Fig. 3. (This scenario may be compared with another possibility for charged stripes shown in Fig. 2.) Such stripe orders are associated with charge excitations so that they are gapped due to long-range Coulomb interaction [45]. The stripes compete with superconductivity and give rise to local $8a_0$ spin periodicity and $4a_0$ charge periodicity. The corresponding gapped incommensurate spin fluctuations differ from the gapless spin-density-wave (SDW) excitations in typical magnetically ordered systems.

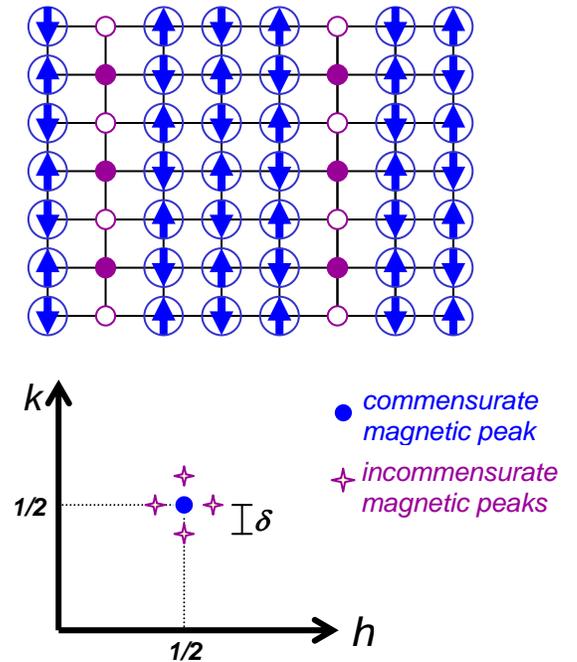

Fig. 3: Schematic illustration of the stripe phase scenario in the $CuO_2$ plane that could give rise to the magnetic diffraction pattern from neutron scattering in the reciprocal space shown in the lower panel. The charge is largely localized in the shaded channels, with a charge density of $+e$ per two sites represented by the alternating solid and open circles. The stripe is an antiphase boundary for the AFM phase illustrated by the blue arrows. For ideal AFM insulating phase, the static magnetic Bragg diffraction is represented by the blue dot at a wave vector $Q = (h,k)2\pi/a_0 = (½,½)2\pi/a_0$, where $a_0$ is the lattice constant in the $CuO_2$ plane. Upon doping with a hole concentration $\delta$, each static Bragg peak is replaced by 4 broadened incommensurate dynamic peaks, indicating spin fluctuations.

The conjecture of a QCP near the optimal doping is motivated by the non-Fermi liquid (NFL) behavior in the normal-state of optimally and under-doped regimes and the Fermi liquid (FL) behavior in the overdoped regime of the p-type cuprates [49]. For the QCP scenario to be relevant to cuprate superconductivity, several criteria must be satisfied. First, a universal broken symmetry at the QCP must be identified and established among all families of cuprate superconductors. Second, how quantum fluctuations associated with the ground state QCP can survive up to the pseudogap temperature to yield non-Fermi liquid behavior must be explained. Third, how the hypothetical QCP near optimal doping may be related to the occurrence of superconductivity and the doping dependence of $T_c$ must be established. Unfortunately, there has been no experimental evidence for a universal broken symmetry at the QCP [9,10], neither has there been adequate justification for an extremely wide temperature range for the quantum critical regime. On the other hand, a QCP could exist between the AFM and SC phases or even between the SC and metallic phases in the ground state. We shall return to this issue later.

The SU(2) gauge theory [50,51] suggests that the pseudogap is associated with the formation of spinon pairing at the pseudogap temperature $T^* \gg T_c$, and that superconductivity occurs at $T_c$ due to the condensation of holons [50,51]. While the model can account for the decreasing $T^*$ and the non-monotonic $T_c$ behavior with increasing hole doping, no apparent spin-charge separation has been detected experimentally, and the premise of $d_{x^2-y^2}$-wave pairing is incompatible with the finding of $s$-wave pairing in a number of n-type cuprates [9,10,17,20].

The BEC scenario, or more precisely, the BEC-BCS crossover theory [46,47], assumes preformed pairs in the pseudogap regime ($T_c < T < T^*$), effective Bose-Einstein condensation (BEC) of these pairs at $T_c$ [46,47], and the Leggett ground state with a large and yet finite attractive coupling $g$ [53]. This theory differs from other phase fluctuation scenarios [42,43] in that the excitations involve both pair excitations and collective modes of the superconducting order parameter, and the excitation gap $\Delta$ is related to the superconducting gap $\Delta_{sc}$ and the pseudogap $\Delta_{pg}$ through a relation $\Delta^2 = \Delta_{sc}^2 + \Delta_{pg}^2$, with the doping dependence of $\Delta_{sc}$ and $\Delta_{pg}$ qualitatively consistent with that of $T_c$ and $T^*$ [46,47]. However, the occurrence of BEC-BCS crossover requires very strong attractive interaction [53], which seems difficult to attain in these doped Mott insulators with strong Coulomb repulsion. Moreover, the generic background AFM correlation in the superconducting state of the cuprates cannot be naturally accounted for in the BEC-BCS crossover scenario.

The SO(5) non-linear σ-model [5,52] assumes Cooper pairing occurs at a mean-field temperature comparable to the magnetic coupling $J$, and describes the competing antiferromagnetic (AFM) and $d_{x^2-y^2}$-wave superconducting (SC) phases by introducing a five-dimensional "superspin" order parameter that approximately commutes with the $t$-$J$ Hamiltonian [31,39,42,54-56] of the cuprates in the long-wavelength limit. Depending on the chemical potential (which is related to the doping level), the ground state can be AFM, SC, or a mixed state of coexisting AFM and SC order, known as a "spin-bag phase" [57]. In addition, collective modes consistent with empirical observation in the AFM and SC states can be derived from solving for the Goldstone modes of the Hamiltonian [5]. However, the model is inherently incompatible with $s$-wave pairing symmetry. Hence, the SO(5) scenario alone is not applicable to some of the n-type cuprates that exhibit $s$-wave pairing symmetry and absence of gapped excitations [9-11,17].

Other theoretical approaches include the mean-field consideration of competing orders in the ground state of the doped Mott antiferromagnet by studying the effects of increasing doping levels and varying strengths of the Coulomb interaction, plus other secondary effects (such as next-nearest neighbor interaction, bi-layer interaction, additional correlation in the orbital- or spin-degree of freedom, etc.) and external magnetic fields on the $t$-$J$ model Hamiltonian [4,58]. These approximations could lead to the occurrence of a QCP within the superconducting state under special conditions [4], as well as other ground states besides the AFM and $d_{x^2-y^2}$-wave SC phases, including stripes and charge density-waves (CDW) [42,59], spin density-waves (SDW)

[60,61], the DDW particle-hole condensate, [4,29,62,63], and ($d_{x^2-y^2}+id_{xy}$) or ($d_{x^2-y^2}+is$)-pairing SC state [4,59].

Finally, there are other conjectures based on more exotic magnetism-driven mechanisms similar to that in fractional quantum Hall effect, such as the anyon superconductivity [64,65]. However, the existence of such mechanism would imply a global broken T-symmetry in the superconducting state of all cuprates, which contradicts most experimental results [9,15,16,66-69].

## 2.2. Debates over the Pairing Symmetry and its Microscopic Implication

The pairing symmetry of cuprate superconductors has been a heavily debated issue over the years [9-22]. Establishment of the symmetry is important because the detailed momentum ($k$) dependence of the order parameter has important implications on the underlying pairing mechanism. For instance, pairing mechanism based on antiferromagnetic spin fluctuations [39,40] would require a superconducting gap with $d_{x^2-y^2}$ pairing symmetry, whereas conjectures based on anion superconductivity [64,65] would favor ($d_{x^2-y^2}+id_{xy}$) pairing symmetry with a complex order parameter that breaks the global T-symmetry. It has also been suggested that the possible existence of a secondary pairing component could be better revealed at surfaces of the cuprates because of the suppression of the dominating $d_{x^2-y^2}$ pairing channel at surfaces due to a surface current within a sheath on the order of the superconducting coherence length [70]. However, such conjecture is inconsistent with the majority of experiments [9,15,16,66-69], except elusive reports from either bulk measurements of cuprate thin films covered with heterogeneous materials [71] or phase sensitive measurements based on scanning SQUID microprobe technique that revealed complex order parameter directly associated with local impurities [72]. Indeed, it has been shown that interface disorder and the degree of interface transparency can drastically affect the tunneling characteristics [73,74], particularly for the zero-bias conductance peak (ZBCP) [21,22] associated with quasiparticle tunneling along the nodal direction of a $d_{x^2-y^2}$-wave superconductor, which would have shown distinct splitting from one peak into two peaks if broken T-symmetry occurred.

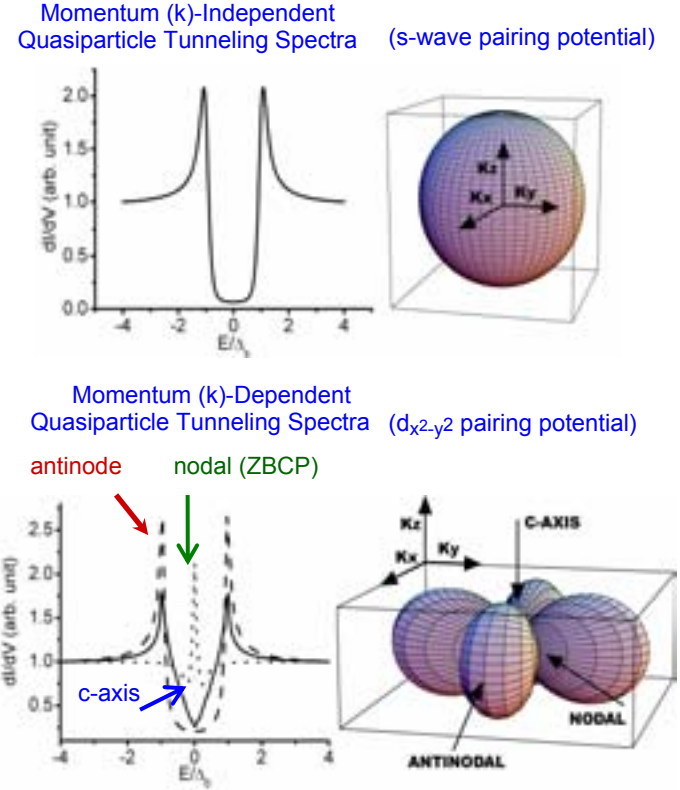

*Fig.4: Comparison of s-wave and $d_{x^2-y^2}$-wave pairing potentials in the momentum (k) space and the corresponding quasiparticles tunneling spectra for quasiparticles momentum along various principal axes. For more details, see Refs. [9,10,15,16,21,22].*

Based on group theory consideration, the pairing channels of a singlet superconductor with a square-lattice symmetry must be consistent with even orbital quantum numbers (such as $s$, $d$, $g$ ... for $\ell$ = 0, 2, 4 ... or their linear combinations). Given the quasi-two dimensional nature of most cuprates and the strong on-site Coulomb repulsion, it is feasible that the pairing symmetry is predominantly associated with the $d$-channel so as to minimize the on-site Coulomb repulsion and to accommodate the quasi-two dimensional nature at the price of a higher kinetic energy. Schematic comparison of the $s$-wave and $d_{x^2-y^2}$-wave pairing potentials and the corresponding quasiparticle tunneling spectra are shown in Fig. 4. Indeed, overwhelming experimental evidences [13-16] are consistent with predominantly $d_{x^2-y^2}$ pairing symmetry (> 95%) for all p-type cuprates in the undoped and optimally doped regimes, and representative directional quasiparticle tunneling spectra of the YBCO system are shown in Fig. 5 for different doping levels.

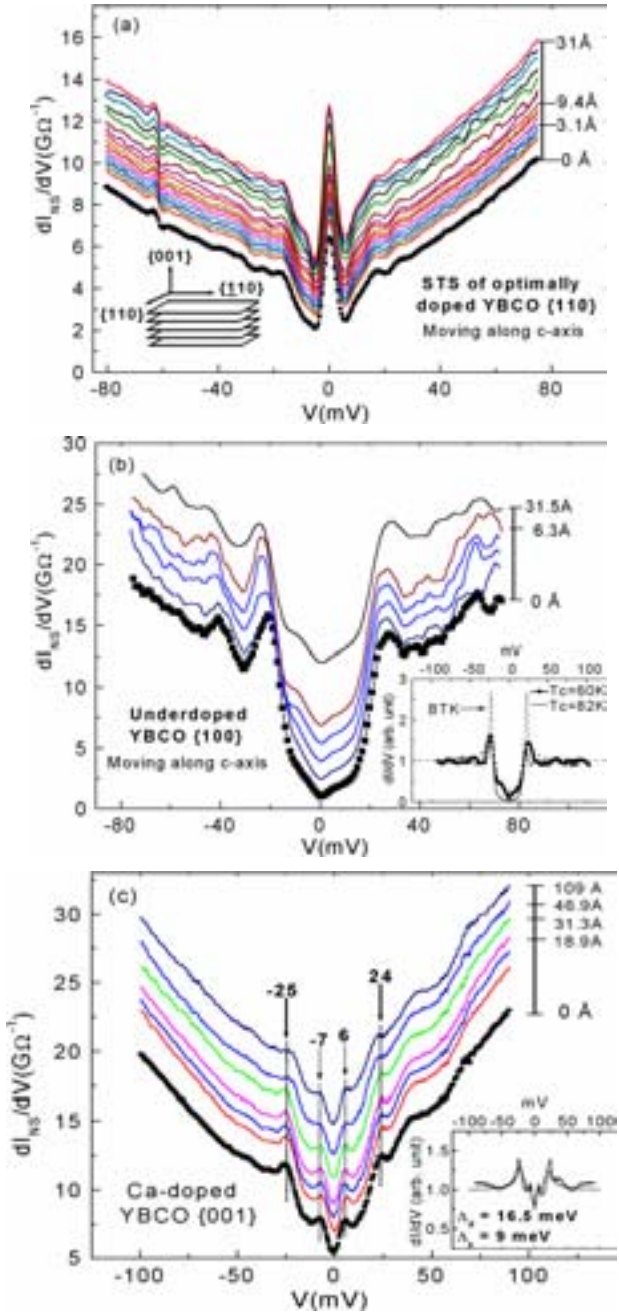

*Fig.5: Differential conductance ($dI_{NS}/dV$) vs. bias voltage (V) tunneling spectra for YBCO at 4.2 K, showing long-range homogeneity [9,15]: **(a)** Optimally doped ($T_c$ = 92.9±0.1K), with quasiparticle momentum **k** || {110} and scanning along {001}. **(b)** Underdoped ($T_c$ = 60.0±1.5 K) with **k** || {100} and scanning along {001}. **Inset:** Comparison of the normalized {100} spectra of underdoped single crystals with $T_c$ = 82 K and 60 K. **(c)** Ca-doped (overdoped) ($T_c$ = 78.0±2.0 K) with **k** || {001} and scanning along {100}. **Inset:** normalized spectrum of a curve in the main panel and the fitting curve with a pairing potential given by $\Delta_k = \Delta_d \cos 2\theta_k + \Delta_s$, where $\Delta_d$ = 18 meV and $\Delta_s$ = 7 meV. (See Refs. [9,15] for more details).*

In contrast, the situation associated with the pairing symmetry of n-type cuprates is far more complicated. Earlier tunneling measurements of the single-crystalline one-layer n-type cuprates have shown results consistent with s-wave pairing symmetry [17], whereas later experiments based on phase-sensitive studies [18] and microwave surface impedance measurements [19] of one-layer n-type cuprates suggest $d_{x^2-y^2}$-wave pairing symmetry. More recently, further studies of the one-layer cuprates suggest doping dependent pairing symmetry [20]. This unsettling issue is in part due to the difficulties in making high-quality n-type cuprates without disordered interstitial oxygen and also the coexistence of magnetism of the rare earth elements and superconductivity of the $CuO_2$ planes [7]. The apparent non-universal pairing symmetry in the one-layer n-type cuprates together with recent finding of s-wave pairing symmetry in the simplest form of cuprate superconductors [9,10], known as the infinite-layer n-type cuprates $Sr_{1-x}Ln_xCuO_2$ (where Ln = La, Gd, Sm, see Fig. 6), strongly suggest that the pairing symmetry in the cuprates may be the consequence of competing orders rather than a sufficient condition for the occurrence of cuprate superconductivity.

## 3. NEW EXPERIMENTAL DEVELOPMENT

In this section we review some of the recent experimental developments that provide important new information for the microscopic descriptions of cuprate superconductivity. Special emphasis will be placed on the issues of competing orders, non-universal pairing symmetry, different low-energy excitations and response to quantum impurities among the p-type and n-type cuprates, and possible physical origin of the pseudogap phenomenon.

### 3.1. Quantum Impurities in P-Type Cuprate Superconductors

Magnetic quantum impurities are known to suppress conventional superconductivity, and the detailed effects have been a topic of great research interest over the years [75-79]. In contrast, non-magnetic impurities in the dilute limit are found to have negligible effects on conventional superconductivity [80]. However, recent findings of strong effects of spinless quantum impurities on p-type cuprate superconductors [15,81-92] have rekindled active investigation on the effects of quantum impurities on superconductivity.

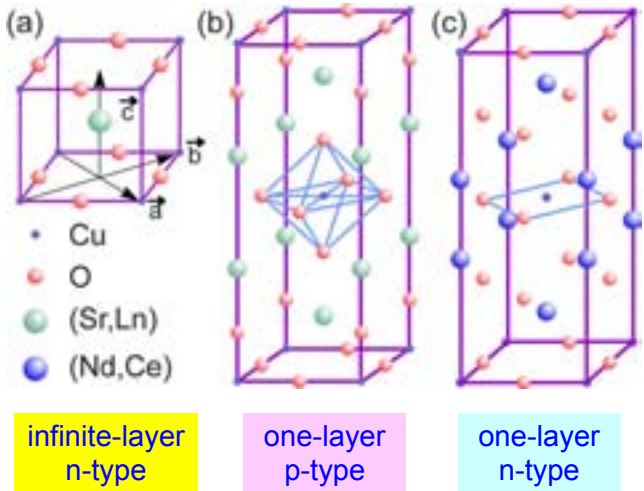

Fig.6: *Comparison of the crystalline structures of representative p-type and n-type cuprates. We note the absence of apical oxygen in all n-type cuprates, in contrast to the existence of $CuO_6$ octahedron or its variations in all p-type cuprates. Moreover, the infinite-layer system differs from all other cuprates in that there is no excess block of charge reservoir between consecutive $CuO_2$ planes.*

Generally speaking, the effects of quantum impurities on superconductivity depend on the pairing symmetry and the existence of magnetic correlation in cuprate superconductors [93-101]. For instance, Fermionic nodal quasiparticles in the cuprates with either $d_{x^2-y^2}$ or $(d_{x^2-y^2}+s)$ pairing symmetry can interact strongly with the quantum impurities in the $CuO_2$ planes and incur significant suppression of superconductivity regardless of the spin configuration of the impurity [93-97], in stark contrast to the insensitivity to spinless impurities in conventional *s*-wave superconductors [80]. Moreover, the spatial evolution of the quasiparticle spectra near quantum impurities would differ significantly if a small component of complex order parameter existed in the cuprate [98]. For instance, should the pairing symmetry contain a complex component such as $(d_{x^2-y^2}+id_{xy})$ that broke the T-symmetry, the quasiparticle spectrum at a spinless impurity site would reveal two resonant scattering peaks at energies of equal magnitude but opposite signs in the electron-like and hole-like quasiparticle branches [98]. In contrast, for either $d_{x^2-y^2}$ or $(d_{x^2-y^2}+s)$ pairing symmetry [15,92], only one resonant scattering peak at the impurity site is expected for large potential scattering strength [93-97]. In addition, the existence of nearest-neighbor $Cu^{2+}$-$Cu^{2+}$ antiferromagnetic coupling in the superconducting state of the cuprates can result in an unusual Kondo-like behavior near a spinless impurity [82,84,88-91,100,101] due to induced spin-1/2 moments on the neighboring $Cu^{2+}$-ions that surround the Cu-site substituted with a spinless ion such as $Zn^{2+}$, $Mg^{2+}$, $Al^{3+}$ and $Li^+$ [82,84,88-91], as schematically illustrated in Fig. 7.

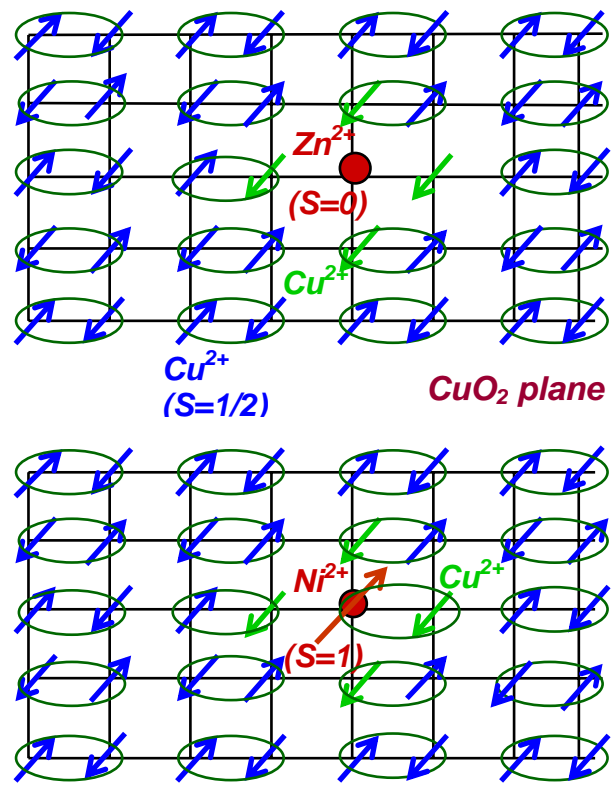

Fig. 7: *Effects of quantum impurities on p-type cuprate superconductors in the underdoped limit.* <u>Upper panel</u>: *Induced magnetic moments on the neighboring $Cu^{2+}$ sites surrounding a spinless impurity (such as $Zn^{2+}$, $Mg^{2+}$, $Al^{3+}$, and $Li^+$ with S = 0) in p-type cuprates.* <u>Lower panel</u>: *A localized $Ni^{2+}$ impurity coexisting with the background AFM coupling in the $CuO_2$ plane.*

Empirically, the Kondo-like behavior associated with isolated spinless impurities in p-type cuprates has been confirmed from the nuclear magnetic resonance (NMR) and inelastic neutron scattering experiments, and the spinless impurities are found to have more significant effects on broadening the NMR linewidth, damping the collective magnetic excitations and reducing the superfluid density than magnetic impurities such as $Ni^{2+}$ with $S = 1$ [82,84,88-91]. On the other hand, both types of impurities exhibit similar global effects on suppressing $T_c$, increasing the microwave surface resistance in the superconducting state and increasing the normal state resistivity [81-102]. The stronger suppression of superconductivity due to spinless impurities in $d$-wave cuprates can be attributed to the slower spatial relaxation of spin polarization near the spinless impurities than that near the $S = 1$ impurities due to the delocalized spatial distribution of the induced moments in the former [99-101], as illustrated in Fig. 7.

Detailed spatial evolution of the quasiparticle tunneling spectra near these quantum impurities in the cuprates can provide useful information for the pairing state of the cuprates, and has recently been investigated in impurity-substituted $Bi_2Sr_2CaCu_2O_{8+\delta}$ (Bi-2212) [92,102] and $YBa_2Cu_3O_{7-\delta}$ (YBCO) [10] systems using low-temperature scanning tunneling microscopy (STM). While in principle both the potential scattering and the Kondo effect contribute to the quasiparticle spectra near spinless impurities, which of the two contributions may be dominant depends on the doping level [101] and cannot be easily determined because direct probing of the quasiparticle spectra near the quantum impurities with scanning tunneling spectroscopy (STS) involves not only the density of states in the $CuO_2$ planes of the cuprates but also the tunneling matrix [100,101]. The tunneling matrix depends on the atomic structure of the surface layers and the exact path of the tunneling quasiparticles [100], which is difficult to determine empirically.

Nonetheless, one can still derive useful information from the STM experimental data by the following simplified consideration. If one 1) neglects many-body interactions in the cuprates, 2) limits the effect of quantum impurities to perturbative and one-band approximation without solving for the spatially varying pairing potential self-consistently [101], and 3) disregards the interaction among impurities, one can describe the effect of quantum impurities with the Hamiltonian $H = H_{BCS} + H_{imp}$. Here $H_{BCS}$ denotes the $d_{x^2-y^2}$-wave BCS Hamiltonian [101] that contains the normal (diagonal) one-band single-particle eigen-energy and anomalous (off-diagonal) $d_{x^2-y^2}$-wave pairing potential $\Delta_k$ (= $\Delta_d \cos 2\theta_k$, $\theta_k$ being the angle relative to the anti-node of the order parameter in the momentum space) of the unperturbed host, and $H_{imp} = H_{pot} + H_{mag}$ denotes the impurity perturbation due to both the localized potential scattering term $H_{pot}$ (= $U \sum_\sigma c_\sigma^\dagger c_\sigma$; $U$: the on-site Coulomb scattering potential) and the Kondo-like magnetic exchange interaction term $H_{mag}$ (= $\sum_R J_R S \cdot \sigma_R$) between the spins of the conduction carriers on the $R$ sites ($\sigma_R$) and those of the localized magnetic moments ($S$).

The above Hamiltonian can be used to obtain the quasiparticle spectra associated with quantum impurities in $d$-wave superconductors by means of Green's function techniques. If one further neglects contributions from the tunneling matrix, one obtains a single resonant energy at $\Omega_0$ on the impurity site in either pure potential scattering limit for a point impurity or pure magnetic scattering limit for four induced moments associated with one spinless impurity [93-97]. For pure potential scattering, one has [94,95]:

$$|\Omega_0/\Delta_d| \approx [(\pi/2 \cot\delta_0 / \ln(8/\pi \cot\delta_0)], \qquad (1)$$

where $\delta_0$ is the impurity-induced phase shift in the quasiparticle wavefunction of a $d_{x^2-y^2}$-wave superconductor, and $\delta_0 \to (\pi/2)$ in the strong potential scattering (unitary) limit. Moreover, the intensity of the resonant scattering is expected to decay rapidly within approximately one Fermi wavelength, and the spatial evolution of the quasiparticles spectra under a given bias voltage V = ($\Omega_0/e$) should reveal 4-fold symmetry of the underlying lattice. Indeed, the spatially resolved STS studies of spinless impurities in optimally doped YBCO and Bi-2212 systems are in reasonable agreement with theoretical predictions for $d_{x^2-y^2}$-wave superconductors [15,92], although whether potential scattering or Kondo effect may be more important has not been determined conclusively. Representative tunneling spectra associated with either $Zn^{2+}$ or $Mg^{2+}$ impurities in YBCO are illustrated in Fig. 8. On the other hand, for magnetic impurities with both contributions from $H_{pot}$ and $H_{mag}$, there are two spin-polarized impurity states at energies $\pm\Omega_{1,2}$ [96]:

$$|\Omega_{1,2}/\Delta_d| = 1/[2N_F(U\pm W)\ln|8N_F(U\pm W)|], \quad (2)$$

where $N_F$ denotes the density of states at the Fermi level and $W \equiv J\mathbf{S}\cdot\boldsymbol{\sigma}$ implies that magnetic impurities are isolated and equivalent at all sites. This prediction for magnetic impurities in $d_{x^2-y^2}$-wave superconductors has been verified by STS studies of Ni-substituted Bi-2212 single crystals [102], and the results are in stark contrast to those of magnetic impurities in conventional $s$-wave superconductors [103,104]. In the latter case, the irrelevance of potential scattering yields only one magnetic impurity-induced bound-state energy at $\pm|\Omega_B|$ and $|\Omega_B| < \Delta_0$, where $\Delta_0$ is the $s$-wave pairing potential, and $|\Omega_B|$ is given by [76]:

$$|\Omega_B/\Delta_0| = (\pi/2)JSN_F. \quad (3)$$

Despite overall similarities in their response to quantum impurities, detailed STS studies of the Bi-2212 and YBCO systems still revealed some interesting differences [15,92]. First, the global superconducting gap $\Delta_d$ was suppressed to $(25 \pm 2)$ meV due to non-magnetic impurities from $\Delta_d = (29 \pm 1)$ meV in pure YBCO [15], whereas the global effect of Zn on Bi-2212 could not be determined because of the strong spatial variations in the tunneling gap values of Bi-2212 [105,106]. Second, the energy $\omega_{dip}$ associated with the "dip-hump" satellite features (see Fig. 8(a)) also shifted substantially relative to that in pure YBCO, whereas such an effect could not be quantified in Bi-2212. The dip-hump feature has been attributed to quasiparticle damping by the background many-body excitations such as incommensurate spin fluctuations [107,108], triplet particle-particle excitations [5,52] or phonons [109], and the resonant energy of the many-body excitation may be empirically given by $|\Omega_{res}| = |\omega_{dip} - \Delta_d|$. We find that the magnitude of $\Omega_{res}$ in the (Zn,Mg)-YBCO sample decreased significantly to $(7 \pm 1)$ meV from that in the pure YBCO where $|\Omega_{res}| = (17 \pm 1)$ meV. This drastic decrease in $\Omega_{res}$ with the very small impurity concentration (<1%) in our Zn and

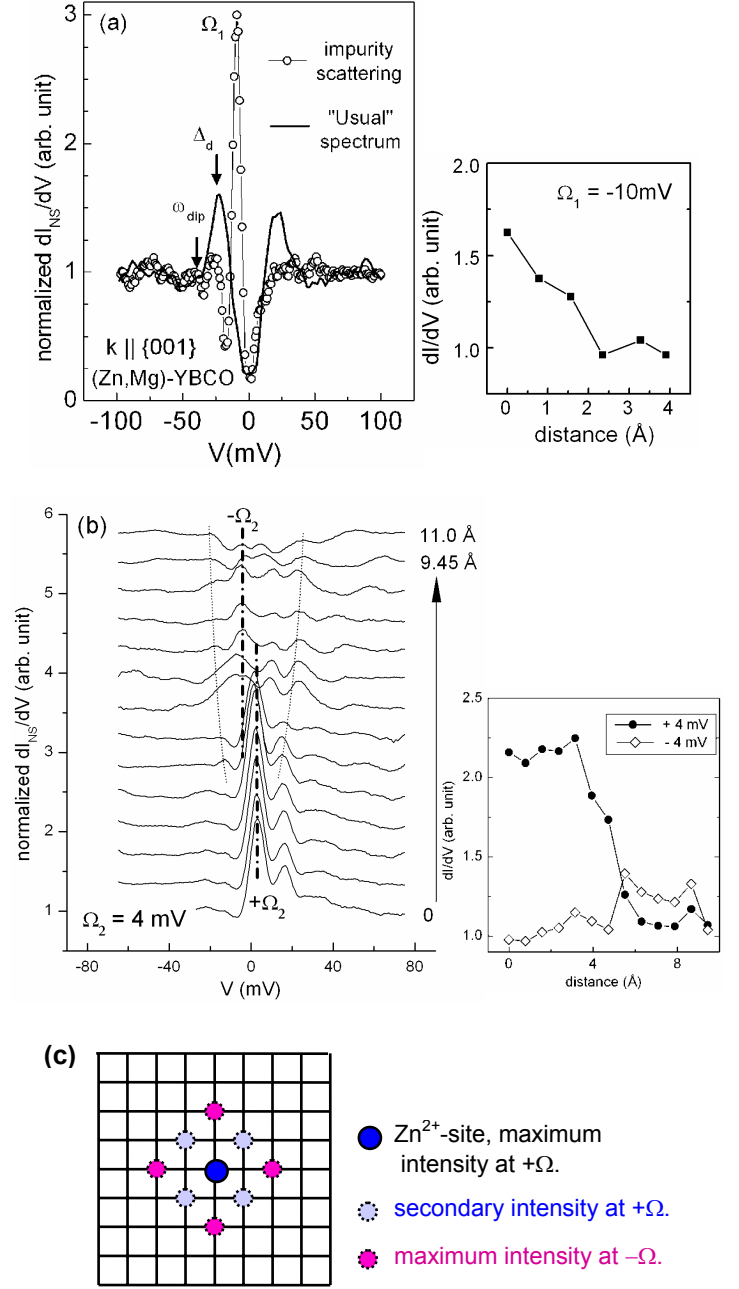

Fig. 8: Normalized c-axis quasiparticle spectra of an optimally doped YBCO near $Zn^{2+}$ or $Mg^{2+}$ impurity at 4.2 K. *(a)* <u>Left panel</u>: An impurity scattering spectrum with a resonant peak at $\Omega_1 \sim -10$ meV and a typical c-axis spectrum far away from impurities. <u>Right panel</u>: Spatial variation of the impurity-induced resonant peak intensity. *(b)* <u>Left panel</u>: Representative spectra revealing spatial variations in the quasiparticle spectra along the Cu-O bonding direction from an impurity with a maximum scattering at $\Omega_2 \sim +4$ meV. We note the alternating resonant peak energies between + 4 meV and − 4 meV and the particle-hole

*asymmetry in the degrees of suppression of the superconducting coherence peaks. Right panel: Spatial variation of the impurity-induced resonant peak intensity at $\pm\Omega_2$. (c) Theoretical predictions [101] for the spatial variations of the impurity scattering intensity at resonant energies $\pm\Omega$ on the $CuO_2$ plane.*

Mg-substituted YBCO has clearly ruled out phonons as the relevant many-body excitations to the satellite features [9,15]. On the other hand, the induced moments due to spinless impurities can suppress the gapped spin fluctuations in the $CuO_2$ planes by randomizing the AFM spin correlation. Third, details of the local spectral evolution near the impurity site also vary somewhat between the Bi-2212 and YBCO systems [15,92]. For instance, the range of impurity effect is longer (~ 3 nm) in YBCO [15] relative to that in Bi-2212 (~ 1.5 nm) [92]. Moreover, the resonant scattering peak in YBCO appears to alternate between energies of the same magnitude and opposite signs near some of the impurities [113], as exemplified in the left panel of Fig. 8(b). Such spatial variations are expected for both Kondo-like and charge-like impurities in *d*-wave superconductors [101].

The response of p-type cuprates such as YBCO [15] and Bi-2212 systems [92] to quantum impurities is empirically in agreement with a pairing state that is gapless along the $(\pm\pi,\pm\pi)$ momentum directions, regardless of the relative strength of potential scattering and magnetic exchange interaction. Therefore the tunneling spectroscopic studies of pure and impurity-substituted p-type cuprates all suggest that the pairing symmetry of p-type cuprates is consistent with pure $d_{x^2-y^2}$ for tetragonal crystals or $(d_{x^2-y^2}+s)$ for orthorhombic crystals [15,16,69,92,102,113], both symmetries involving nodes in the pairing potential along $(\pm\pi,\pm\pi)$. These studies place an upper bound of less than 5% for any secondary complex pairing component [15,16,69].

### 3.2. Strongly Correlated S-Wave Pairing in the Infinite-Layer N-Type Cuprates

As discussed earlier in Section 2.2, the pairing symmetry in the n-type cuprates appears to be non-universal and doping dependent [9,10,17-20]. In particular, the simplest form of cuprate superconductors [114,115], known as the infinite-layer n-type cuprates $Sr_{1-x}Ln_xCuO_2$ (where Ln = La, Gd, Sm, see Fig. 6), reveal strong spectroscopic evidences for a pure *s*-wave pairing symmetry, although the pairing state still differ significantly from conventional weak-coupling characteristics [9,10,113]. In this subsection, we summarize the experimental evidence for strongly correlated *s*-wave superconductivity in the infinite-layer system. The specific aspects for consideration include: 1) momentum-independent quasiparticle tunneling spectra and the absence of satellite features, 2) conventional response to quantum impurities, and 3) consistency of magnetic impurity-induced bound states with the Shiba states for *s*-wave superconductors. We also critically examine the relevance of pseudogap to cuprate superconductivity, particularly given the absence of pseudogap in all n-type cuprates under zero magnetic field [9-12].

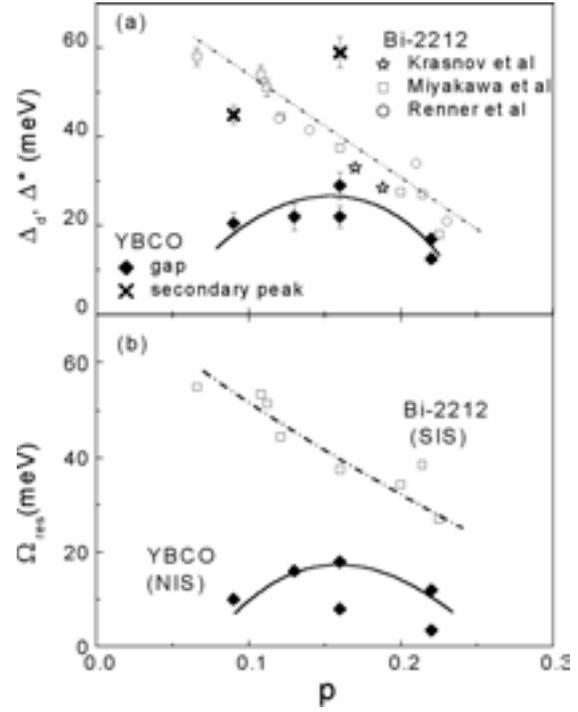

*Fig. 9: (a) The tunneling gap $\Delta_d(p)$ of YBCO, as determined from STS measurements, is compared with the measured gap $\Delta^*(p)$ in Bi-2212 from various techniques, including direct measurements on mesa structures (Krasnov et al. [110]), point contact and S-I-S break-junction measurements (Miyakawa et al. [111]), and STS studies (Renner et al. [112]). The doping level p, except for the optimally doped (Zn,Mg)-YBCO, is estimated by means of an empirical formula $T_c = T_{c,max} [1-82.6(p-0.16)^2]$, with $T_{c,max} = 93.0$ K for the optimally doped YBCO. The global value of $\Delta_d$ in the optimally doped (Zn,Mg)-YBCO is reduced relative to that of pure YBCO. (b) Comparison of $\Omega_{res}(p)$ and $\Omega_2(p)$ for YBCO and Bi-2212. Note the resemblance of $\Omega_{res}(p)$ to $\Delta_d(p)$, and the significant suppression of $\Omega_{res}$ due to spinless impurities.*

Concerning the issue of pairing symmetry, it is feasible that the pairing symmetry of p-type cuprates favors $d_{x^2-y^2}$-wave over *s*-wave in order to minimize the on-site Coulomb repulsion and the orbital potential energy while maintaining the quasi-two dimensionality, because the presence of apical oxygen in p-type cuprates (see Fig. 6) lifts the degeneracy of $d_{x^2-y^2}$ and $d_{3z^2-r^2}$ orbitals in favor of $d_{x^2-y^2}$-orbital for holes, as discussed earlier. On the other hand, the absence of apical oxygen in the n-type cuprates retains the degeneracy of $d_{x^2-y^2}$ and $d_{3z^2-r^2}$, thus favoring a more three-dimensional pairing. In the case of one-layer n-type cuprates, the large separation between consecutive $CuO_2$ planes could still favor a $d_{x^2-y^2}$-wave pairing symmetry that preserves the quasi-two dimensionality, although the exact pairing symmetry in a specific cuprate depends on the subtle balance of competing energy scales as a function of electron doping and also on the degree of oxygen disorder in the interstitial sites between $CuO_2$ planes.

In contrast, the infinite-layer n-type cuprates such as $Sr_{0.9}La_{0.1}CuO_2$ differ from other cuprates in a number of ways. First, the infinite-layer system contains only one metallic monolayer of Sr or La rather than a large charge reservoir as in other cuprates between consecutive $CuO_2$ planes. Second, the c-axis superconducting coherence length ($\xi_c \sim 0.53$ nm) is longer than the c-axis lattice constant ($c_0$) [116,117], in contrast to the typical condition of $\xi_c << c_0$ in most other cuprates. Hence, the infinite-layer system is expected to reveal more three-dimensional characteristics. Third, Knight-shift experiments [11] have revealed that the carrier density of the optimally doped $Sr_{0.9}La_{0.1}CuO_2$ at the Fermi level is significantly smaller than that in typical p-type cuprates, being ~ 25% that of optimally doped $YBa_2Cu_3O_{7-\delta}$. These atypical characteristics of the infinite-layer system are suggestive of a tendency towards more isotropic pairing symmetry and strong electronic correlation.

Despite their importance to better understanding of cuprate superconductivity, the infinite-layer n-type cuprates are very difficult to synthesize, and the lack of single-phased compounds with high volume fraction of superconductivity has hindered research progress until a recent breakthrough [116,117]. Using high-pressure (~ 4 GPa) and high-temperature (~ 1000 °C) annealing conditions, Jung *et al.* have been able to achieve single-phased $Sr_{0.9}Ln_{0.1}CuO_2$ compounds with nearly ~ 100% superconducting volume [116]. The availability of these high-quality infinite-layer cuprates has enabled our STS studies of the quasiparticle tunneling spectra and the pairing symmetry, yielding some curious characteristics that defy widely accepted notions derived from p-type cuprate superconductors [9,10].

First, the quasiparticle tunneling spectra and the superconducting energy gap $\Delta$ appear to be momentum-independent, as manifested by spectra taken on more than 300 randomly oriented single crystalline grains [9,10] and exemplified in Fig. 10. Second, the ratio of $(2\Delta/k_BT_c) \sim 7$ for $T_c$ = 43 K is much larger than the BCS ratio (~ 3.5) for weak coupling *s*-wave superconductors. Third, no discernible satellite features exist in the quasiparticle spectra, in sharp contrast to those of all p-type cuprates, as manifested by the two insets of in Fig. 10 for normalized tunneling spectra taken on optimally doped YBCO and $Sr_{0.9}La_{0.1}CuO_2$ (La-112). It is worth noting that in the context of *t-J* or Hubbard model, the satellite features are strictly associated with *d*-wave pairing symmetry [5,52,107,108]. Fourth, the tunneling gap features completely vanish above $T_c$, suggesting the absence of a pseudogap [9,10], which is also independently verified by NMR experiments [11]. Fifth, the global response of the system is fundamentally different from that in p-type cuprates, being insensitive to non-magnetic impurities such as Zn up to 3% and extremely susceptible to magnetic impurities such as Ni so that superconductivity becomes completely suppressed with <3% Ni substitution [9,10,117], as manifested in Fig. 11.

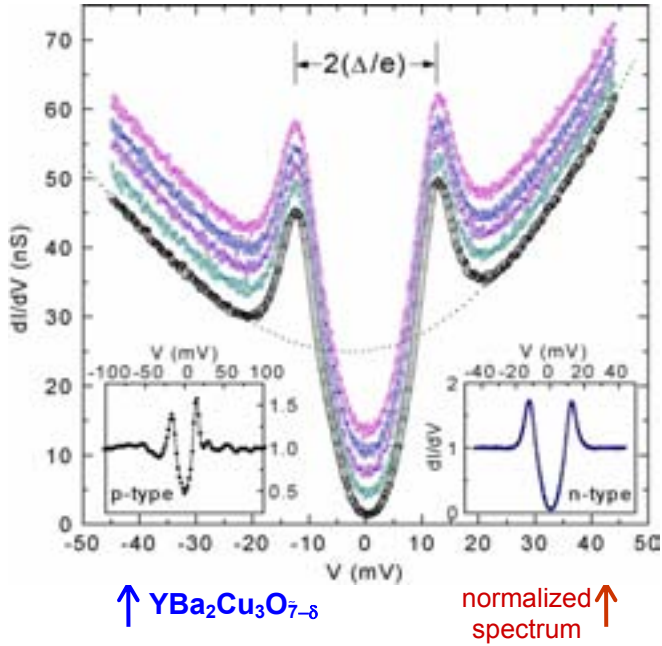

↑ YBa$_2$Cu$_3$O$_{7-\delta}$   normalized↑ spectrum

*Fig.10: <u>Main Panel</u>: Representative quasiparticle spectra taken on pure $Sr_{0.9}La_{0.1}CuO_2$ and at 4.2 K, showing momentum-independent spectral characteristics and vanishing DOS at zero bias. <u>Right inset</u>: Normalized spectrum relative to the background conductance shown as the dotted line in the main panel. We note the absence of satellite features for $|V|>\Delta/e$ and excess DOS at $0 < |V| < \Delta/e$. <u>Left inset</u>: Normalized c-axis tunneling spectrum of a YBCO single crystal, showing significant satellite features at high energies, in contrast to the spectrum of $Sr_{0.9}La_{0.1}CuO_2$.*

As described in the previous subsection, cuprate superconductors with $d_{x^2-y^2}$ pairing symmetry are strongly affected by both magnetic and non-magnetic quantum impurities in the CuO$_2$ planes. On the other hand, superconductors with s-wave pairing symmetry are insensitive to a small concentration of non-magnetic impurities due to the fully gapped Fermi surface and therefore limited interaction with the low-energy excitations at low temperatures [80]. Thus, the global response of the infinite-layer system to quantum impurities is indeed consistent with s-wave pairing symmetry. Assuming the validity of the Abrikosov-Gor'kov theory [75], we estimate $J \sim$ 0.3 eV [113] for $Sr_{0.9}La_{0.1}CuO_2$ with a critical magnetic impurity concentration $x_c \sim 3\%$. This exchange energy is comparable to but somewhat larger than the $Cu^{2+}$-$Cu^{2+}$ antiferromagnetic coupling constant.

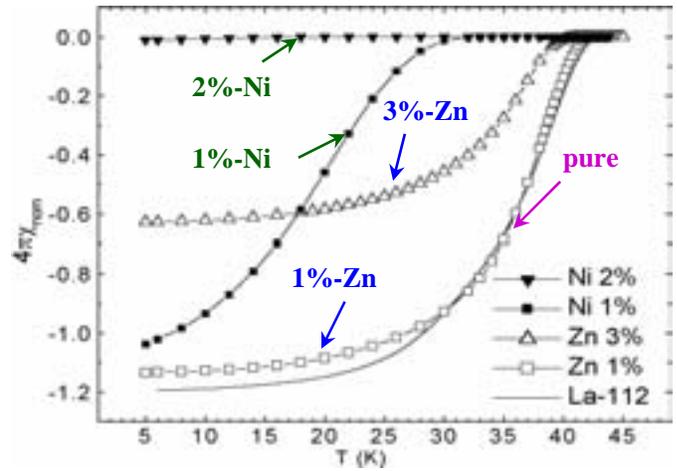

*Fig.11: Bulk magnetic susceptibility data of pure $Sr_{0.9}La_{0.1}CuO_2$ (La-112) and those with small concentrations of impurities. While the superconducting volume consistently decreases with increasing impurities, the superconducting transition temperature ($T_c \sim$ 43 K for pure La-112) reveals little dependence on spinless Zn-impurity substitution up to 3%, and drastic decrease of with magnetic Ni-impurity substitution [9,117].*

Although the momentum-independent pairing potential $\Delta$ is supportive of a fully gapped Fermi surface, details of the spectral characteristics appear different from those of weak-coupling isotropic s-wave superconductors. To examine whether the discrepancy may be the simple result of anisotropic pair potential, we have performed the generalized BTK analysis [16,21,22] and concluded that any anisotropy exceeding 8% should have yielded resolvable momentum-dependent variations in the quasiparticle spectra [113], as exemplified in Fig. 12. So what may have been the physical origin for the excess sub-gap quasiparticle DOS (see the right inset of Fig. 10) in the infinite-layer cuprates despite a momentum-independent energy gap and the vanishing quasiparticle DOS at the zero bias that rules out disorder-induced effects? The answer may lie in the unusual low-energy excitations in n-type cuprates. That is, the deviation from the spectra of conventional s-wave superconductors may be attributed to the coupling of thermally induced quasiparticles to the background SDW. As stated before, the low-energy spin excitations in n-type superconducting cuprates are gapless SDW according to neutron scattering experiments [118]. These low-energy excitations are absent in conventional *s*-wave superconductors so that the latter reveal little sub-gap DOS at low temperatures. The presence of SDW in n-type cuprates may also weaken Cooper pairing, thus yielding generally lower $T_c$ in n-type cuprates.

DOS associated with the infinite-layer cuprates, the absence of discernible satellite features is also noteworthy, as manifested in the inset of Fig. 10. We have described in previous sections that the satellite features in p-type cuprates can be attributed to quasiparticle damping by gapped spin excitations along the Cu-O bonding direction [107,108]. Hence, the absence of such satellite features is consistent with the absence of gapped incommensurate spin fluctuations and *s*-wave superconductivity in the infinite-layer system.

Further verification for the pairing symmetry can be made via studying the response of the superconductor to magnetic and non-magnetic impurities. As shown in Fig. 11, the bulk response of the infinite-layer system to quantum impurities differs substantially from that of p-type cuprates [117] and resembles that of conventional *s*-wave superconductors. Moreover, detailed investigation of the local quasiparticle spectra reveals additional support for the *s*-wave pairing symmetry in the infinite-layer system. That is, the tunneling gap value of optimally doped La-112 with 1% Zn impurities remains comparable to that of pure La-112 with no apparent spatial variations, although excess sub-gap quasiparticle density of states exists due to disorder [9,10,113]. In contrast, significant particle-hole asymmetry is induced in the quasiparticle tunneling spectra of the La-112 sample with 1% Ni impurities [9,10,113], as shown in Fig. 13(a). The long range impurity-induced density of states in Fig. 13(b) is also consistent with the extended Shiba states [76] for magnetic impurity bands in *s*-wave superconductors, and only one bound-state energy $|\Omega_B| \sim 5$ meV can be identified [113], in contrast to the local quasiparticle spectra near magnetic impurities in *d*-wave superconductors [95,102] where strong quasiparticle spectral variations near a magnetic impurity and two different impurity-induced resonant energies are observed. It is interesting to note that the exchange interaction *J* derived from Eq. (3) with empirical values of $|\Omega_B|$, $\Delta_0$ and $N_F$ is also consistent with the estimate using Abrikosov-Gor'kov theory [75] with a critical magnetic concentration $x \sim 0.3$ [113]. Hence, all spectral characteristics of the Ni-substituted La-112 sample are consistent with those of a strongly correlated *s*-wave pairing superconductor.

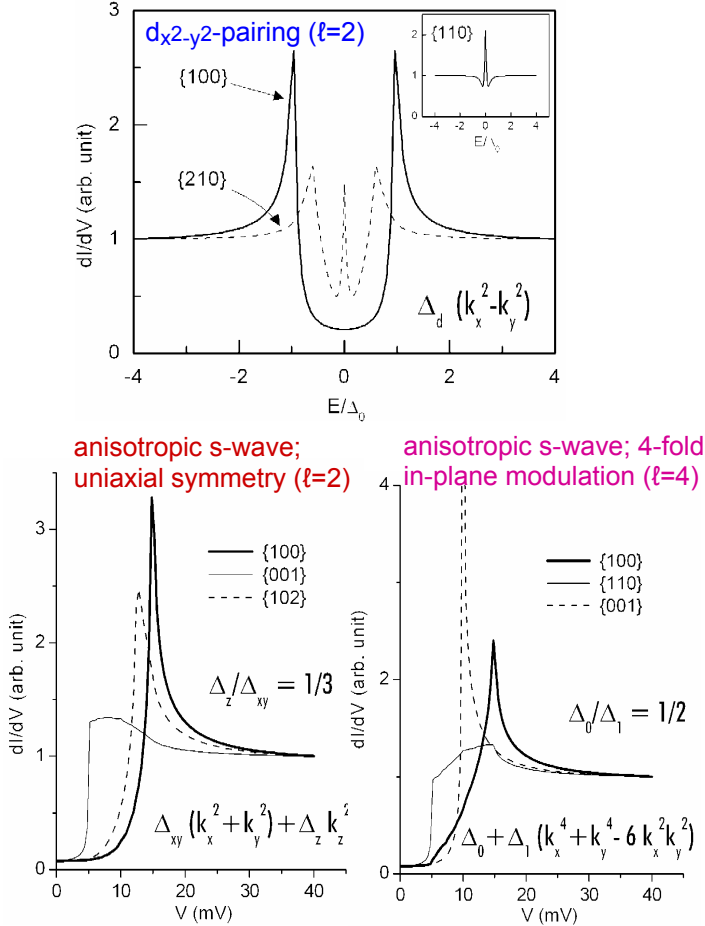

*Fig.12: Calculated quasiparticle tunneling spectra for various pairing symmetries with anisotropic pairing potentials $\Delta_k$ as given. Upper panel corresponds to different quasiparticle tunneling momenta into a pure d-wave superconductor. The lower left panel corresponds to those of an anisotropic s-wave pairing potential with uniaxial symmetry, and the lower right panel depicts the spectra of anisotropic s-pairing with 4-fold in-plane modulations [113].*

Besides the momentum-independent spectral characteristics and excess sub-gap quasiparticle

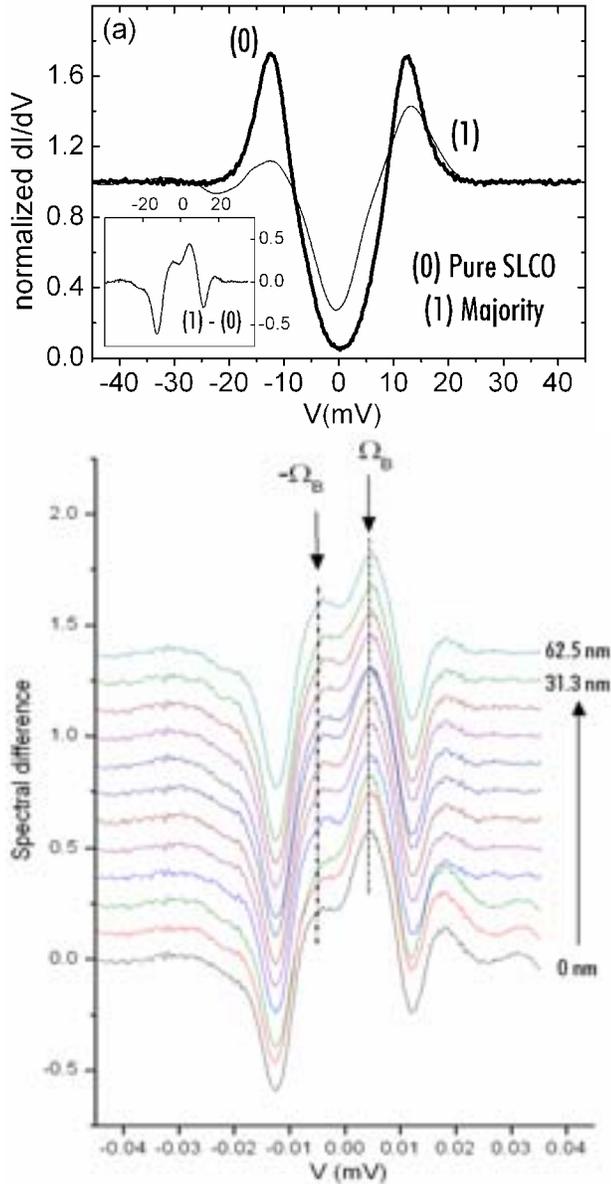

*Fig.13: **(a)** Comparison of the quasiparticle spectra taken on pure $Sr_{0.9}La_{0.1}CuO_2$ and on $Sr_{0.9}La_{0.1}(Cu_{0.99}Ni_{0.01})O_2$ at 4.2 K. The inset illustrates the spectral difference of the two spectra in the main panel, which corresponds to Ni-impurity contributions. **(b)** Long-range spatial extension of the impurity spectral contribution. The spectra have been shifted vertically in the graph for clarity. These spectra appear to be quite homogeneous over long range within each grain, consistent with the Shiba states for impurity bands. Two asymmetric bound-state energies in the electron-like and hole-like branches are visible at $|\Omega_B| \sim 5$ meV, corresponding to $J \sim 0.3$ eV for $\Delta_0 = 13$ meV. (See Ref. [113] for more details).*

In addition to *s*-wave pairing symmetry, the optimally doped La-112 system exhibits complete absence of pseudogap above $T_c$ from both tunneling studies [9,10] (Fig. 14) and the Knight shift measurements [11]. Recent tunneling spectroscopic studies of the one-layer n-type cuprates $Pr_{2-x}Ce_xCuO_{4-y}$ [12] also reveal no pseudogap phenomenon above $T_c$ for a wide range of doping levels in zero field, while the application of high magnetic fields at $T \ll T_c$ results in an effective pseudogap at $T^* < T_c$ for several underdoped samples, with $T^*$ decreasing with increasing electron doping and vanishing at the optimal doping level. Hence, the pseudogap phenomenon is obviously not a precursor for superconductivity in n-type cuprates.

### 3.3. Remark on the Origin of the Pseudogap

Regarding the physical origin of the pseudogap phenomenon, we conjecture that in p-type cuprates the decreasing zero-field $T^*$ with increasing hole-doping may be correlated with gapped spin excitations such as the incommensurate spin fluctuations [24-28] or triplet pair excitations [5,52], so that the decreasing spin stiffness with increasing doping naturally yields a decreasing $T^*$. The gapped spin excitations imply spin-singlet states exist between $T_c$ and $T^*$, which are effectively preformed pairs with physical properties different from those of conventional Fermi liquid. In contrast, the presence of gapless SDW excitations in n-type cuprates may imply that spin-singlet pairs cannot exist above $T_c$ because of the incompatibility of SDW with spinless singlet pairs once the superconducting gap vanishes. On the other hand, the application of a large magnetic field competes with the background AFM spin correlation, so that the resulting low-energy spin excitations in n-type cuprates could change from gapless SDW to gapped spin-flip processes, thereby yielding an effective pseudogap. Moreover, the energy cost for spin flips under a constant magnetic field is expected to decrease with decreasing spin stiffness, which is consistent with a decreasing field-induced $T^*$ that decreases with the increasing doping level. Thus, our conjecture of the pseudogap being a manifestation of quasiparticle damping by gapped spin excitations in doped cuprates has provided a consistent phenomenology for the following experimental facts: 1) the doping dependence of $T^*$ in p-type cuprates; 2) the non-Fermi liquid behavior in the pseudogap regime of p-type cuprates.; 3) the absence of zero-field pseudogap and the doping dependence of a field-induced pseudogap in n-type cuprates; and 4) the excess sub-gap quasiparticle DOS in n-type cuprates at $T << T_c$.

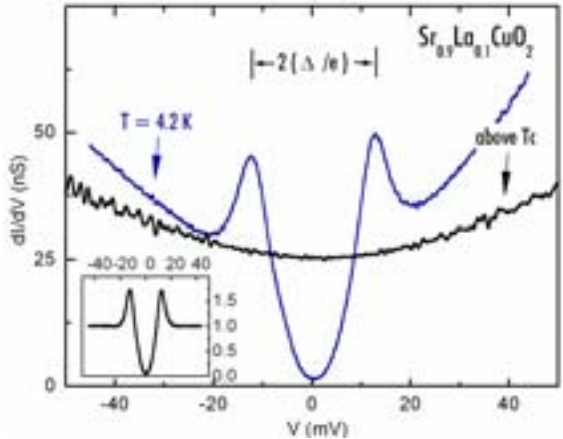

*Fig. 14: Comparison of the tunneling spectra of $Sr_{0.9}La_{0.1}CuO_2$ taken at $T = 4.2$ K (~ 0.1 $T_c$) and at $T >\sim T_c$, showing complete absence of any pseudogap above $T_c$.*

### 3.4. Competing Orders in the P-Type Cuprate Superconductors

In the presence of competing orders, a specific order parameter can prevail if other orders are suppressed by external variables. For instance, doping dependence of the resistive state properties of various p-type cuprates has been investigated by applying large magnetic fields at low temperatures, with a metal-to-insulator crossover behavior found at a doping level well below the optimal doping [119], implying no QCP near the optimal doping. On the other hand, neutron scattering studies of the vortex state of $La_{2-x}Sr_xCuO_4$ (x = 0.16 [120] and 0.12 [121]) and $La_{1.875}Ba_{0.125-x}Sr_xCuO_4$ [122] have revealed that the AFM spin ordering within the vortex core is enhanced to the extent comparable to that in the normal state, while the spin correlation extends over a spatial range substantially longer than the vortex-vortex separation [120-122]. Moreover, the spin correlation exhibits $8a_0$-periodicity, suggesting a related $4a_0$-periodicity for the charge [120]. This interesting observation associated with the vortex cores of p-type cuprates is further corroborated by the STS studies of an optimally doped Bi-2212 system, where directly observation of $4a_0 \times 4a_0$ "checker-board" low-energy (< 12 meV) charge structures within the vortex cores are made [123]. The spectroscopic findings were initially interpreted as the manifestation of competing AFM and superconductivity [58,60]. That is, the AFM spin order and the accompanying charge order is presumably enhanced due to the suppression of superconductivity in the vortex core and in the regions surrounding the vortex cores due to the presence of induced supercurrents [58,60]. However, further STS studies of the Bi-2212 system in the absence of field [124] also reveal similar checker-board patterns for large areas of the sample, prompting reevaluation of the original interpretation [124]. By performing Fourier analyses on the energy-dependent spatial conductance modulations of the spectra, dispersion relations consistent with those derived from angular resolved photoemission spectroscopy (ARPES) [125,126] are found. This finding suggests that the zero-field conductance modulations in STS data of Bi-2212 are primarily the result of interferences due to elastic scattering of quasiparticles between states of equivalent momenta on the Fermi surface of the superconductor [124]. This simple explanation has effectively ruled out the possibility of charge stripes as a competing order in the Bi-2212 system, because the presence of charge stripes would have resulted in momentum-independent Fourier spectra, in contrast to the strongly dispersive spectra [124]. As for the excess checker-board like conductance modulations within the vortex cores under the application of large dc magnetic fields [123], it is yet to be verified whether a similar scenario, based on quasiparticle interferences due to elastic scattering between equivalent states on the field-driven normal-state Fermi surface, can account for the large magnitude of conductance modulations inside the vortex core [123,124]. It is worth noting that the quasiparticle interference scenario [124] cannot easily account for either the magnetic field-induced enhancement of AFM spin correlations [120-122] or the metal-to-insulator transition [119] in the La-Sr(Ba)-Cu-O system. Hence, competing orders of AFM and superconductivity may still be relevant when one considers the field-induced effects on cuprate superconductivity.

Another seemingly controversial issue regarding the spatial variation of the superconducting order parameter in different cuprates [15,105,106] can also be understood in the context of competing orders. That is, it has been noted recently from STS studies that nano-scale variations exist in the tunneling gap of the Bi-2212 system [105,106], with nano-scale regions of sharp superconducting coherence peaks embedded in a "less superconducting" background of pseudogap-like broadened tunneling peaks in the spectra. These nano-scale regions are comparable in size while the density of these regions increases linearly with hole-doping level [105], and the spectra eventually become spatially homogeneous for strongly overdoped samples [127]. On the other hand, no such nano-scale variations can be found in the YBCO system, as manifested by STS studies of a wide doping range of YBCO samples that revealed the long-range (~ 100 nm) spatially homogeneous spectral characteristics [9,15], and by NMR studies of similar systems [128]. The different behavior between YBCO and Bi-2212 can be understood as two types of doped Mott insulators that respond differently to the doping level, similar to the different response of type-I and type-II superconductors to an applied magnetic field [129]. More specifically, consider two competing phases A and B in a strongly correlated electronic system, as schematically illustrated in Fig. 15. Depending on the magnitude of the effective inertia and interaction potential in the Hamiltonian of the physical system, different behavior as a function of the chemical potential ($\mu$) can exist [5,130,131]. If Phases A and B are separated by a first-order critical point or a critical line as depicted in Fig. 15(a), nano-scale phase separations can occur for $\mu \sim \mu_c$. On the other hand, if Phases A and B can coexist over a range of doping levels, as depicted in Fig. 15(b), the sample would reveal long range phase homogeneity for the intermediate doping range. Finally, glassy behavior would occur in the crossover regime if disorder dominates between Phases A and B, as shown in Fig. 15(c). Thus, the nano-scale order-parameter variations in Bi-2212 may be associated with the phase diagram in Fig. 15(a) while the long-range spatially homogeneous order parameter in YBCO may be related to the phase diagram in Fig. 15(b).

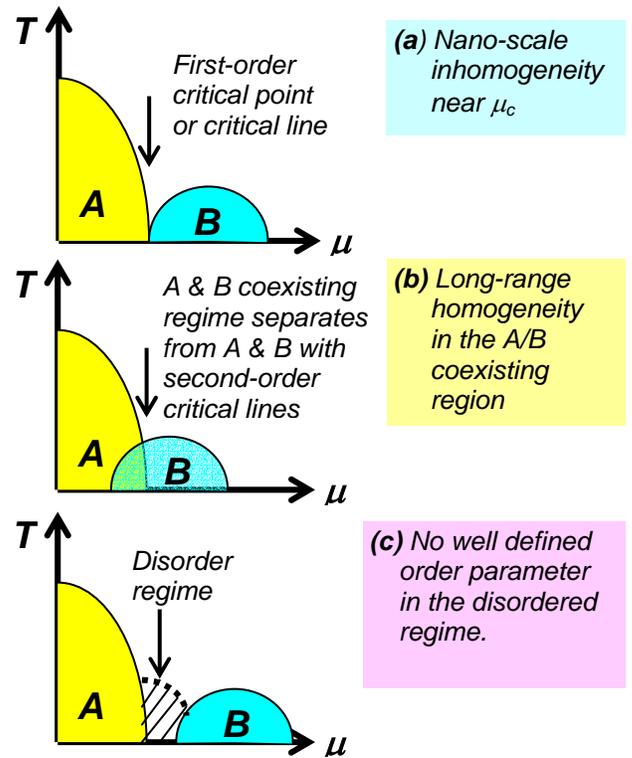

Fig. 15: Possible temperature (T) vs. chemical potential phase ($\mu$) diagrams for two competing phases A and B. Different behavior depends on the energy scales of competing terms in the Hamiltonian.

The question is: what may be the relevant competing phases in YBCO and Bi-2212, and what may be the differences between YBCO and Bi-2212 that give rise to varying spatial homogeneity in the superconducting order parameter? We speculate that the competing orders in the p-type cuprates may be the pseudogap phase and the superconducting state. The former corresponds to a phase with incommensurate gapped spin excitations and the latter is an effective spin liquid. On the difference(s) between YBCO and Bi-2212 that may be responsible for determining the magnitude and sign of the domain wall energy between the competing phases, we suspect that the large anisotropy in Bi-2212 (particularly in underdoped samples) versus the stronger three dimensional coupling in YBCO may contribute to the occurrence of nano-scale phase separations in the former. This issue awaits further theoretical investigation. We also remark that the formation of nano-scale phase separations is by no means a necessary condition for superconductors with short coherence lengths, as some might have naively assumed. In fact, different ground states as a function of the chemical potential have also been observed in the perovskite manganites $Ln_{1-x}M_xMnO_3$ (Ln: La, Nd, Pr, M: Ca, Sr, Ba), which are strongly correlated electronic systems showing colossal negative magnetoresistance (CMR) effects [130-132]. Depending on the doping level and the chemical composition, the competing phases of ferromagnetism (FM) and AFM in the manganites can result in nano-scale inhomogeneity in the magnetic order parameter, as empirically manifested by STM imaging [132] and theoretically verified via numerical calculations [130,131].

### 3.5. Unusual Effects of Spin-Polarized Quasi-Particles on Cuprate Superconductivity

Given the relevance of AFM correlation to cuprate superconductivity and the drastic effects of quantum impurities on the physical properties of the pairing state, one may consider an interesting scenario of injecting spin-polarized quasiparticles into the cuprates and investigating the relaxation process of the polarized spin currents. In addition, by comparing spin-injection experimental results with data derived from simple quasiparticle injection, one can obtain useful information for the spin and charge transport mechanisms in cuprate superconductors and investigate the possibility of spin-charge separation [35,137]. Indeed, such experiments have been conducted by fabricating layered structures of perovskite ferromagnet (F), non-magnetic metal (N), insulator (I) and cuprate superconductor (S), with ferromagnetic CMR manganites $La_{0.7}Ca_{0.3}MnO_3$ (LCMO) and $La_{0.7}Sr_{0.3}MnO_3$ (LSMO) chosen for the F-layer, non-magnetic $LaNiO_3$ (LNO) for the N-layer, $SrTiO_3$ (STO) or yttrium-stabilized zirconium (YSZ) for the I-layer, and $YBa_2Cu_3O_{7-\delta}$ (YBCO) for the S-layer. Such heterostructures can be grown epitaxially on perovskite substrates (e.g. $LaAlO_3$), yielding high-quality interfaces without degradation to the constituent layers [35,133-136], thereby ensuring minimum interface quasiparticle spin scattering and maximum spin polarization for the injection currents. The spin polarized currents can be obtained by passing electrical currents through half-metallic [138-140] ferromagnetic manganites before forcing them into the superconducting layer, as illustrated in Fig. 16.

Systematic studies of the critical current density ($J_c$) of the superconducting layer in perovskite F-I-S and N-I-S heterostructures as a function of the injection current density ($J_{inj}$) and temperature (T) have been made on samples with a range of thicknesses for the F-, N- and I-layers [35,135]. In addition, STS studies of the quasiparticle DOS under finite $J_{inj}$ have been performed on the YBCO layer [136]. These measurements reveal much stronger effects of spin-polarized quasiparticles than those of simple quasiparticles on cuprate superconductors [35,135,136]. Further analyses of the data indicate that conventional theory of non-equilibrium superconductivity [141,142] is not applicable to the spin-injection phenomena in cuprate superconductors [35], and that the spin relaxation mechanism in the cuprate appears anisotropic, with unusually slow in-plane spin relaxation, probably due to the long-range disruptive effects of spin polarized currents on the background AFM correlation [35]. On the other hand, the c-axis spin relaxation is much faster, with a characteristic relaxation time comparable to that associated with the spin-orbit interaction [35]. These findings underscore the importance of AFM correlation to the integrity of cuprate superconductivity, and are also supportive of quasiparticles rather than solitons (e.g. spinons and holons) as the relevant low-energy excitations in the superconducting state of the cuprates [35].

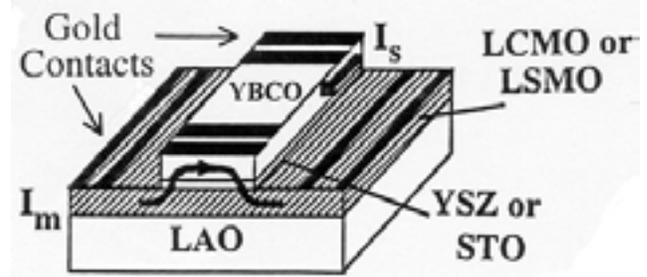

$I_s$: currents applied directly to YBCO.
$I_m$: currents applied directly to the manganite.

*Fig. 16: Schematic illustration for spin injection experiments in perovskite F-I-S samples. For more details, see Refs. [35,135].*

### 3.6. Strong Phase Fluctuation Effects & Novel Vortex Phases and Dynamics

Being doped Mott insulators with layered structures, the order parameter of cuprate superconductors reveals reduced phase stiffness and strong fluctuation effects, particularly in the underdoped p-type cuprates [42,143]. The fluctuation effects are manifested in the magnetic, electrical and thermal transport properties. In particular, novel vortex phases and dynamics in the mixed state of cuprate superconductors have yielded rich experimental phenomena and new theoretical understanding [143-145]. The most significant difference between the vortex state of cuprate superconductors and that of conventional type-II superconductors is the existence of a vortex-liquid state [143] intermediate between a vortex-solid state and the normal state of the former. In particular, the vortex dynamics of the cuprates appears to be very sensitive to disorder and anisotropy, so that the equilibrium vortex-solid state varies from the Abrikosov lattice to Bragg glass [146] or vortex glass [147,148] under random point disorder, to Bose glass under parallel columnar defects or twin boundaries along the crystalline c-axis [149,150], to splayed glass under canted columnar defects [151,152], or pinned Josephson vortices for magnetic fields parallel to the $CuO_2$ planes [145,153,154]. Moreover, different types of phase transitions can exist within the vortex-solid phase [155,156], between the vortex-solid and vortex-liquid phases [157-159], and within the vortex-liquid state [160], depending on the disorder and anisotropy of the cuprate. In addition, anomalous sign reversal in the Hall conductivity with varying temperature and magnetic field is found in the vortex-liquid state of both p-type and n-type cuprate superconductors [161-164]. These rich phenomena are believed to be the direct consequence of strong phase fluctuations in the cuprates, and the physical origin for the strong fluctuations is obviously tied to the microscopic theory of the cuprates.

Besides experimental manifestation of strong fluctuation effects on vortex dynamics in the superconducting state, other noteworthy findings associated with fluctuation effects in the normal state include non-vanishing superfluid density above $T_c$, as obtained from complex conductivity of Bi-2212 [165], and enhanced Nernst effect associated with vortex excitations in $La_{2-x}Sr_2CuO_4$ and $Bi_2Sr_2LaCu_2O_{8-\delta}$ above the upper critical field $H_{c2}(T)$ and $T_c$ [166]. However, the range of fluctuations in the normal state appears significantly different between the zero-field [165] and high-field [166] experiments, the former being substantially smaller than the latter. Moreover, while the zero-field complex conductivity data [165] above $T_c$ can be understood in terms of phase fluctuations of the superconducting order parameter, the enhanced Nernst effect in the normal state [165] cannot be explained with simple phase fluctuations alone. These results suggest that the application of large magnetic fields not only suppresses the superconducting order parameter but also influences the background spin correlation in the cuprates. Thus, the physical properties of p-type cuprates in the pseudogap regime appear to differ from conventional fluctuation conductivity. A full description for the unconventional properties may have to involve microscopic consideration for the spin correlation and pair excitations.

## 4. FURTHER DISCUSSION & OUTLOOK

After reviewing a wide variety of experimental information associated with both p-type and n-type cuprates, it is clear that no obvious particle-hole symmetry exists in these doped Mott insulators, so that the simple approach of a one-band Hubbard model cannot provide a universal account for all experimental findings. In particular, it appears that only two commonalities can be identified among all families of cuprates. One is the strong electronic correlation and the other is the AFM spin correlation in the $CuO_2$ planes [9,10]. A number of important phenomena previously deemed as essential to cuprate superconductivity are in fact not universal, including the $d_{x^2-y^2}$ pairing symmetry, the pseudogap phenomena and incommensurate spin fluctuations. These latest experimental developments have thus imposed stringent constraints on existing theories.

Can a sensible physical picture emerge from all experimental facts associated with both p-type and n-type cuprates while simultaneously reconcile a number of seemingly conflicting observations? Empirically, we note that an important difference between p-type and n-type cuprates is in the low-energy spin excitations, although both systems retain short-range AFM $Cu^{2+}$-$Cu^{2+}$ spin correlation in their superconducting state [6,25-28,118]. For arbitrary doping levels, incommensurate spin fluctuations could occur along the Cu-O bonding direction of p-type cuprate superconductors. These spin fluctuations are gapped and are therefore suppressed in the ground state. The anisotropic spin excitation gap, quasi-two dimensionality and the tendency to minimize on-site Coulomb repulsion in p-type cuprates could conspire to yield the lowest ground state energy under pair wavefunctions with $d_{x^2-y^2}$-symmetry. Moreover, for a given doping level, the incommensurate spin excitation gap of p-type cuprates is always larger than or comparable to the superconducting gap [25-28], implying that singlet pairing of carriers can exist in the $CuO_2$ planes at temperatures below the incommensurate spin excitation gap, and that the relevant "mean-field" energy scale is $\Omega_{res}$ rather than the AFM exchange energy $J$. This scenario is consistent with the presence of a pseudogap and the existence of singlet pairs in the pseudogap regime ($T_c < T < T^*$) of the p-type cuprates. Moreover, the fluctuation effects in the pseudogap regime are primarily associated with the charge rather than the spin degrees of freedom.

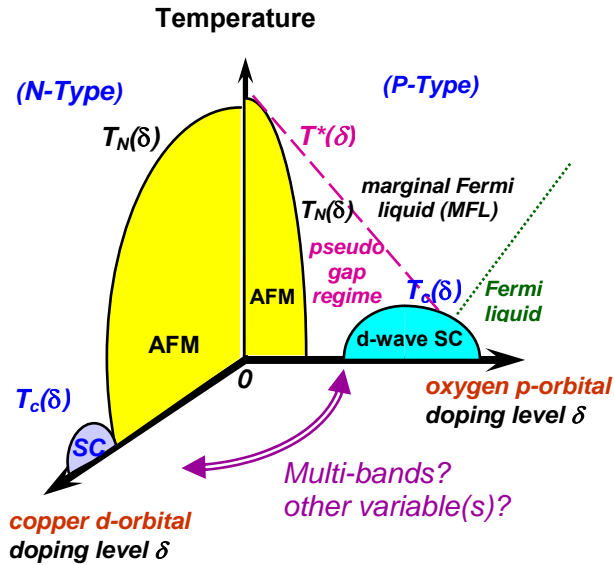

Fig. 17: *An attempt to unify the phase diagrams of p-type and n-type cuprates may require the consideration of additional degrees of freedom.*

On the other hand, the low-energy spin excitations in n-type cuprates are gapless SDW [118]. The presence of SDW in the superconducting state could hinder singlet pairing because of the tendency of misaligned spin orientation for pairs over a finite spatial distance, thus yielding generally lower $T_c$ values in one-layer n-type cuprate superconductors relative to one-layer p-type cuprates. The absence of gapped incommensurate spin excitations in n-type cuprates is also consistent with the absence of pseudogap. As for the pairing symmetry, it is conceivable that the combined effects of strong three dimensional electronic coupling in the infinite-layer system (see Fig. 6), the existence of isotropic SDW excitations and the degeneracy of $d_{x^2-y^2}$ and $d_{3z^2-r^2}$-orbitals would favor s-wave pairing symmetry in the ground state. On the other hand, the quasi-two dimensionality energy in the one-layer n-type cuprates may compete with the aforementioned energy scales so that the overall energy difference between s- and $d_{x^2-y^2}$-wave pairing is small and strongly dependent on the doping level and oxygen disorder.

Despite the consistency of the above scenario with most experimental observation, it provides no microscopic description for the Cooper pairing in the $CuO_2$ planes. While it is clear that AFM spin correlation plays an important role in the pair formation and pseudogap phenomena, the link to unifying the phase diagrams of p-type and n-type cuprates is yet to be identified. Meanwhile, most phenomenology such as the stripe scenario or the DDW model can be regarded as special cases of competing orders rather than a sufficient condition for cuprate superconductivity. Thus, the primary theoretical challenge is to address the inadequacy of one-band Hubbard model and to examine whether multi-band approximation or inclusion of other variable(s) may be necessary in the quest of unifying the phenomenology of all cuprates, as schematically illustrated in Fig. 17. Ultimately, the development of an adequate microscopic theory for this strongly correlated electronic system must prescribe an effective attractive pairing interaction among carriers that suffer strong on-site Coulomb repulsion. The effective attraction may result from unique pair wavefunctions with optimized orbital and spin degrees of freedom that minimize the Coulomb repulsion, and the resulting effective attraction is likely to be only moderate compared with the bare Coulomb energy. In fact, the possibility of a moderate-to-small effective attraction may explain why certain physical properties associated with the cuprates can be reasonably modeled with the BCS approximation, although it is highly probable that the pairing mechanism for cuprate superconductivity is fundamentally different from conventional electron-phonon mediated BCS superconductivity, and may unavoidably involve magnetism.

## 5. CONCLUDING REMARKS

The discovery and subsequent intense research of high-temperature superconducting cuprates have revolutionized our understanding of superconductivity and strongly correlated electronic materials. We have reviewed in this article some of the recent experimental developments and the status of various theoretical scenarios, and have suggested that many interesting experimental findings can be understood in terms of competing orders. On the other hand, the apparent differences among hole-doped (p-type) and electron-doped (n-type) cuprates are indicative particle-hole asymmetry and of the inadequacy of considering the cuprates in terms of a one-band Hubbard model. It is conjectured that different forms of low-energy spin excitations in the cuprates, i.e. gapped incommensurate spin fluctuations in the p-type and gapless SDW in the n-type, may play an important role in determining the ground state and low-energy excitation spectra of the corresponding cuprate superconductors. In particular, the pseudogap phenomenon may be associated with the gapped incommensurate spin excitations, and therefore is absence in n-type cuprates. The pairing symmetry is also non-universal and appears to be a consequence of competing orders. The only ubiquitous properties among all cuprates are the strong electronic correlation and AFM spin interaction in the $CuO_2$ planes. Future research challenge will require the convergence of empirical facts and the development of a microscopic theory that unifies all experimental observation and provides an effective attractive interaction for pair formation in the $CuO_2$ planes of the cuprates.

## 6. ACKNOWLEDGEMENT

The author is pleased to acknowledge collaborations with Ching-Tzu Chen and Chu-Chen Fu at Caltech, Dr. Richard P. Vasquez and Dr. L. Douglas Bell at the Jet Propulsion Laboratory, Prof. Jochen Mannhart and his research group at the University of Augsburg in Germany, Dr. Satuko Tajima and her research group at the Superconductivity Research Laboratory of ISTEC in Japan, and Prof. Sung-Ik Lee and his research group in Pohang University of Science and Technology in Korea. Useful discussions with Prof. Dung-Hai Lee, Prof. S. Uchida, Prof. Elbio Dagotto and Dr. Xiao Hu are gratefully acknowledged. The research at the California Institute of Technology (Caltech) has been supported by the National Science Foundation, Grant # DMR-0103045.